\documentstyle[twocolumn,prl,aps]{revtex}
\large
\begin{document}
\draft
\twocolumn[
\hsize\textwidth\columnwidth\hsize\csname@twocolumnfalse\endcsname

\title
      {
      Crystal-like high frequency phonons in the amorphous
      phases of solid water
      }
\author{
	H.~Schober$^{1}$,
        M.M.~Koza$^{2}$,
	A.~T\"olle$^{2}$,
	C.~Masciovecchio$^{3}$,
        F.~Sette$^{4}$, and
	F.~Fujara$^{5}$.
       }
\address{
         $^1$
	 Institut Laue-Langevin, F-38042 Grenoble Cedex, France.\\
	 $^2$
	 Fachbereich Physik, Universit\"at Dortmund,
	 D-44221 Dortmund, Germany\\
	 $^3$
         Sincrotrone Trieste, Area Science Park,
	 I-34017, Trieste, Italy. \\
	 $^4$
         European Synchrotron Radiation Facility,
         F-38043 Grenoble Cedex, France.\\
	 $^5$
         Technische Universit\"at Darmstadt, Hochschulstr 6, D-64289
         Darmstadt, Germany. \\
	         }

\date{\today}
\maketitle
\begin{abstract}
The high frequency dynamics of low- ($LDA$) and high-density
amorphous-ice ($HDA$) and of cubic ice ($I_c$) has been measured
by inelastic X-ray Scattering (IXS) in the 1$\div$15 nm$^{-1}$
momentum transfer ($Q$) range. Sharp phonon-like excitations are
observed, and the longitudinal acoustic branch is identified up
to $Q = 8$~nm$^{-1}$ in $LDA$ and $I_c$ and up to 5~nm$^{-1}$ in $HDA$.
The narrow width of these excitations is in sharp contrast with the
broad features observed in all amorphous systems studied so far.
The "crystal-like" behavior of amorphous ices, therefore, implies a
considerable reduction in the number of decay channels available
to sound-like excitations which is assimilated to low local disorder.
\end{abstract}


]
Amorphous polymorphism, i.e.\ the existence of two or more amorphous
states in the phase diagram of a chemical substance, has recently
attracted wide interest from the scientific community
\cite{Poole-Science-1997}.
In water, amorphous polymorphism has received particular attention
as it was associated with a possible phase separation in the deeply
undercooled liquid \cite{Harrington-PRL-1997} - this is suggested
by the idea that the two amorphous water phases may be
identified with the glassy forms of two liquid phases. In systems
like water, the ergodic to non-ergodic transition from the undercooled
liquid to the glass cannot be studied continuously as a function of
temperature due to homogeneous crystallization. The glassy nature of
the amorphous ice phases can, therefore, only be established indirectly,
and this has contributed to raise controversies on the exact nature of the
phase-diagram of water, and on the inter-relations among the different
stable and metastable phases of this molecule \cite{Tse-Nature-1999}.

The growing perception that the glassy state is a very general state of
condensed matter justifies the  many experimental, theoretical and
simulation studies performed on glasses. Among them, there is a growing
interest in the study of the collective dynamical properties at high
frequency, i.e.\ in the wavelength regime approaching the interparticle
distances. Here one expects that either a continuum description, valid
in the hydrodynamic limit, or the ordered-medium (phonon) picture
developed for crystalline materials will fail as a consequence of the
disorder. The deviations can then be used to characterize and better
understand the glassy state. More specifically, there is a number of
dynamic signatures which are associated with the glassy state
\cite{Fischer-PRL-1999}. At very low temperatures two-level states
give rise to tunneling phenomena \cite{tunneling}. At higher temperatures
disorder scattering and anharmonicities play an increasingly
important role leading to excess densities of states (Boson peak) and
relaxational phenomena \cite{relaxational}. Recently, the overdamping of
sound waves
at high frequency was proposed as a further criteria to characterize
the glassy behavior \cite{Sette-Science-1998}.
The study of the collective dynamics
of amorphous ice at wavelengths approaching the distance among water
molecules may contribute, therefore, to the understanding of the
differences between  these two phases, and may help to shed light on
such issues as their relation to stable crystalline phases or on the
existence of two different liquid phases.

In the present letter we present the excitation spectrum of both
amorphous forms of water, i.e.\ high-density ($HDA$) and low-density
amorphous ice ($LDA$), as determined by high resolution inelastic X-ray
scattering (IXS) at momentum transfers ($Q$) comparable to the
inverse of the inter-molecular distance.
The experiment has been performed at the very high energy resolution
ID16 beamline at the European Synchrotron Radiation Facility in Grenoble.
The instrument energy  resolution was set to 1.6\,meV full width half maximum
(FWHM) using 21.748\,keV incident photons. The detection system was made
up of five independent analyzer systems, displaced among each other by a
constant
angular offset corresponding to a $Q$-spacing of 3~nm$^{-1}$.
The $Q$ resolution was set to 0.4~nm$^{-1}$ FWHM.
The energy scans at constant $Q$-transfer took about 150', and each $Q$ point
was obtained by a typical averaging of 3-5 scans.
The transverse dimensions of the beam at the sample were
0.15\,$\times$\,0.3\,mm$^{2}$. Further experimental
details can be found elsewhere \cite{expdet}.

The $HDA$ sample
was obtained by pressurizing hexagonal D$_2$O
ice $I_h$ at 77\,K \cite{Mishima-Nature-1984}
beyond 10\,kbar using the piston-cylinder
apparatus described previously \cite{Koza-PRL-1999}.
Keeping the sample always at liquid nitrogen
temperature, the metastable compound was retrieved
from the cell at ambient pressure in mm-size chunks ($\rho = 1.17 \pm
0.02$\,g/cm$^3$),
and placed into a steel container with two diametrically opposed
openings to allow for the passage of the incident
and scattered X-ray beam. The effective powder sample
thickness along the beam was $\approx$ 15\,mm, matching
well the X-ray photo-absorption length. Once filled,
the container was transferred under a continuous helium
flow from the liquid nitrogen bath onto the cold-finger
of a precooled closed-cycle refrigerator. Above $\approx 90$\,K
$HDA$ transforms with a strongly temperature dependent
rate to $LDA$ ($\rho = 0.94 \pm 0.02$\,g/cm$^3$), which,
in turns, converts itself into cubic ice $I_{\rm c}$ at
$\approx 140$\,K. $HDA$ and $LDA$, after proper annealing,
were both measured at 60\,K, while $I_c$ after annealing
was measured at 80\,K. The similarity of the temperatures
allows a direct comparison of the scattering intensities.

The purity of each phase was checked by measuring the
respective static structure factors, $S(Q)$, reported
in Fig.~1 with a momentum resolution of 0.04 nm$^{-1}$.
Both $HDA$ and $LDA$ samples show the known static
structure factor \cite{Bellissent-JCP-1987} with no signs of
Bragg-peaks and, therefore, they are free of crystalline
ice XII \cite{Koza-Nature-1999}. There is a
pronounced small-angle signal in $HDA$ which disappears
upon annealing to $LDA$. Within our energy resolution,
its origin is purely elastic as it will be seen by the
inelastic scans reported in the following.
As only moderate changes in the elastic small angle signal
were observed with neutrons \cite{Schober-PB-1998}, it is
questionable whether the strong effect observed here is
a real bulk property of the $HDA$ sample or whether its origin
stems from surface scattering in the powder sample.

A selection of inelastic spectra obtained in the
range $Q < 8$\,nm$^{-1}$ are reported in Fig.~2 and Fig.~3.
In both amorphous ice phases and in the cubic crystalline
phase there are well pronounced resonances.
Increasing $Q$ one initially observes a single feature with a
marked dispersion.
At $Q$-values higher than 5~nm$^{-1}$, the spectra become more
complicated by the appearance of additional inelastic features.
As X-rays, like neutrons, couple directly to the longitudinal component of the
density fluctuations, the dispersive excitation observed from the lowest
$Q$-values
can be readily identified with the longitudinal acoustic-like branches.
In $HDA$ these excitations can be clearly observed up to about
5\,nm$^{-1}$, while,
in $LDA$ and $I_{c}$, they can be safely distinguished from the other
feature(s) up
to 10\,nm$^{-1}$.

In order to derive the energy position, $\hbar\Omega(Q)$, and the width,
$\hbar\Gamma(Q)$,
of the excitations, they have been fitted by a Lorenzian function convoluted
with the
experimentally determined instrumental resolution function.
This resolution function corresponds to the central peak at $E = 0$ in
Fig.~2, that,
as discussed previously, is due to the static disorder in the samples and
to small angle scattering
from the sample environment (see Fig.~2 and Ref.\cite{expdet}).
The best fit to the inelastic X-ray scattering data shows that the
excitation energy $\hbar\Omega(Q)$ scales linearly with the
momentum transfer $Q$ in the limit of small momentum transfers.
>From the derived linear relation, $\Omega(Q) = cQ$, it is
possible  to obtain the longitudinal sound velocities
$c_{HDA}= 3550 \pm 50 $\,m/s, $c_{LDA} = 3550 \pm 50 $\,m/s and $c_{I_c}=
3750 \pm 50 $\,m/s, respectively.
The differences among these values are small, and this is in agreement with
the similar Debye level found in these three materials by neutron
scattering experiments \cite{Schober-PB-1998}.
The determination from the best fits of the resonance widths,
$\hbar\Gamma(Q)$, is more
involved, and is complicated by two facts:
{\it i)} the absolute values for $\Gamma(Q)$ are highly correlated with
the background, and
{\it ii)}, particularly in $HDA$, the features appearing with increasing
$Q$ on the low-frequency
side of the acoustic-like peaks interfere with the fit.
Therefore, no systematic dependence of $\Gamma(Q)$ on $Q$ has been
established.
In both $LDA$ and $HDA$ one observes, however, that the resonances become
broader with increasing $Q$-transfer.
In any case care has to be taken in interpreting such broadening
because the line shape of the resonances is unknown.
For a single phonon it could be assumed Lorenzian reflecting the finite
lifetime of the excitation.
However, when taking a powder average the line shape acquires a
non-analytic form even for the crystal due to the anisotropy within
the multidimensional dispersion-sheet.

Despite the difficulty to give full account of the observed broadening in
the measured spectra,
one can deduce from Figs.~2 and 3 the following important observation:
As seen by the naked eye, the width of the resonances remains by far
smaller than the excitation
energy, $\hbar\Omega(Q)$, for all the considered $Q$-values.
This result is very different from the observation made so far in other
glassy systems.
There, in fact, one observes an acoustic phonon-like resonance with
a linear dispersion of $\Omega(Q)$\,vs\,$Q$ and a quadratic dispersion of
$\Gamma(Q)$\,vs\,$Q$ up to a value $Q = Q_m$.
$Q_m$ is defined by the relation $\Omega(Q_m) \approx \Gamma(Q_m)$.
At $Q$ larger than $Q_m$ it is no longer possible to observe
well defined excitations and the inelastic part of the spectrum is, at the
most, a broad and
structureless background.

Let us take the position, $Q_M$, of the first sharp
diffraction peak as an indicator for the extent of
structural correlations in the two amorphous phases.
This allows us to define a pseudo Brillouin zone
\cite{Caprion-PRL-1996} which extends to
$Q_M/2\approx 10$\,nm$^{-1}$ in $HDA$, and
to $Q_M/2\approx 8$\,nm$^{-1}$ in $LDA$.
The quantity $Q_m/Q_M$ has so far always been found smaller than
0.5 \cite{Sette-Science-1998}.
In the two amorphous ice phases studied here,
excitations are very well defined up to $Q$-transfer values approaching
$Q_M$, and, at least
in $LDA$, the longitudinal acoustic-like branch can be identified
very well at least up to $Q = 8$\,nm$^{-1}$.
In Ref.~\cite{Sette-Science-1998}, it has been suggested the possible
existence of a relation
between the value of $Q_m/Q_M$ and the degree of fragility of the
considered glass \cite{Angell}.
The high value of $Q_m/Q_M$, which seems to approach unity in these two
glasses, would
imply that amorphous ices, and especially $LDA$, are the extreme end of
fragile glasses.
This is in contradiction to the view that water passes through an
inflection in the deeply
supercooled region where the liquid behavior changes from extremely fragile
to strong \cite{Ito-Nature-1999,Starr-2000}.

A further observation coming from the analysis of the spectra in Fig.~2 and
3 is that the spectra
of $LDA$, at all the considered $Q$-values, are very similar to those
measured in ice $I_c$ \cite{footnote2}.
In $HDA$ the resemblance is less pronounced \cite{footnote1}.
The similarity of $LDA$ and $I_c$ holds not only for the resonances
assigned to the acoustic mode,
but also for the excitations at $\approx$7\,meV, which appear at higher $Q$
and recall the transverse
dynamics found in ice $I_h$ and liquid water \cite{expdet}.
In fact, as in these systems, the excitations set in around 5\,nm$^{-1}$
and, as seen in Fig.~3,
they become more pronounced beyond $Q_{M}$.
The translational part of the density-of-states for $I_h$, $I_{c}$ and
$LDA$, as obtained from
INS spectra \cite{Schober-PB-1998}, is peaked around 6.5 meV (D$_{2}$O).
It is, therefore to be expected, that the INX spectra show excitations in
this energy region at
low $Q$ due to Umklapp processes.
These processes take place via the Bragg peaks in the crystalline state and
via the static
structure factor in the case of $LDA$ \cite{Buchenau-ZP-1985}.

Despite the high definition of the low $Q$ excitations in the two amorphous
ice phases when compared
to other glasses, one recovers a clear indication of the disordered
character from the evolution of
the inelastic spectrum at larger $Q$ values.
The spectra of $LDA$ and $HDA$ are less structured, lacking the sharp
features observed in ice
$I_{c}$.

To conclude, we reported on an IXS measurement of the $S(Q,E)$ of the two
known phases of
amorphous ice.
This has allowed to show that these two states of the water molecule
possess a surprisingly
crystal-like dynamic response.
In both $HDA$ and $LDA$ the sound wave excitations are well defined.
These experimental findings are in sharp contrast to the results found so
far in other
glasses, glass forming materials, liquids, dense gases and disordered
materials in
general.
In these systems, in fact, an important broadening has always been observed
in the
inelastic part of the dynamic structure factor $S(Q,E)$.
In this $Q$-region the scattering experiment becomes
sensitive to the topological disorder which opens decay channels for
sound excitations in addition to those available in the crystal.
These channels are found practically absent in the case of the amorphous ice
phases indicating a very low degree of local disorder.
Structural results are controversially discussed and to date do not give a
clear
picture of the topology \cite{Bellissent-EPL-1998,Pusztai-PRB-2000}.
Experimentally only the pair correlation functions are directly accessible.
Higher order correlations, among these the orientational correlation function,
must be obtained in an indirect way \cite{Soper-JCP-1994} e.g.\ via the
dynamic response.
We deduce from our data highly intact hydrogen bond networks both in $LDA$
and to
some lesser degree equally in $HDA$.
Although an intact network in itself is no warranty for the absence of
decay channels ---
 e.g.\ an infinite random framework of corner-linked SiO$_{2}$-tetrahedra
can undergo
large phonon-assisted distortions \cite{Trachenko-PRL-1998} --- it seems a
necessary condition.
In $LDA$ this network is perfectly annealed as it is not obtained via a
fast quench from the liquid.
Due to the constraints of the network the number of states the system
can sample on the ps time scale should be small, i.e. there is a
small configurational entropy, a view which is compatible with
thermodynamic data
\cite{Starr-2000}.
$HDA$ is expected to possess a larger configurational entropy, and
 on this basis one can justify that $HDA$ has more "glassy behavior" than
$LDA$.

Apart from structure and bonding the dynamical properties of the water
molecule influence the decay of sound-like excitations both in crystalline
and amorphous ice.
We just want to point to the clear separation of translational and
librational bands which independent of the structural details arises
from the very small moment of inertia of the water molecule.
This separation closes decay channels --- e.g.\ present in SiO$_{2}$
\cite{Taraskin-PRB-1999}--- involving resonances of acoustic-like and
librational modes, and may equally explain the absence of strong
excess intensities in the inelastic neutron scattering data
\cite{Schober-PB-1998}.
Temperature and the concomitant anharmonicities equally have to be
given proper consideration in the discussion \cite{Taraskin-PRB-1999}.
In the end only detailed molecular dynamics calculations on well characterized
ensembles combined with experiments on similar systems will be able to
unambiguously give the reasons for the crystal like dynamic response of
amorphous
ice phases.

We acknowledge A.~Mermet for his help during the IXS measurements,
B.~Gorges and R.~Verbeni for technical support, F.~Sciortino,
C.A.~Angell and G.~Ruocco  for useful discussions.
We also acknowledge the financial support of the Bundesministerium f\"ur
Bildung und Forschung under project number 03{\sc fu}4{\sc dor}5.

\begin{center}
{\bf FIGURE CAPTIONS}
\end{center}

{\footnotesize{
\begin{description}
\item  {FIG. 1 -
Static structure factor of high density ($HDA$) and
low density ($LDA$)  amorphous ice as measured on the inelastic
X-ray beamline prior to the inelastic experiments.
No signs of Bragg peaks are observed, indicating
that both amorphous phases are free of crystalline
ice XII contaminations. Note the strong small-angle signal in $HDA$.
The inset shows the diffraction pattern of cubic ice.
}

\item  {FIG. 2 -
Inelastic X-ray spectra of high density amorphous ($HDA$),
low density amorphous ($LDA$) and crystalline cubic
ice ($I_c$) at the indicated $Q$ values lying mainly in the first
pseudo Brillouin zone.
The dashed line is a fit to the sinal using Lorenzian lineshapes convoluted
with the resolution function (equally indicated).
The solid line represents the inelastic contribution to the total fits.
The numbers in brackets on the left of the elastic
line give the elastic intensities in arbitrary units.
Note the close resemblance
of the sharp inelastic response of both amorphous phases to
the crystalline phase.
}

\item  {FIG. 3 -
Inelastic X-ray spectra of high density amorphous ($HDA$),
low density amorphous ($LDA$) and crystalline cubic
ice ($I_c$) at high $Q$ values in the second pseudo Brillouin zones.
The dashed line gives the resolution function.
The  elastic intensities are given in brackets.
For these high $Q$ values
the excitations become part of a broad intensity distribution
reminiscent of the density-of-states.
}

 \end{description}
}}

\begin{references}

\bibitem{Poole-Science-1997}
		P.H.~Poole T.~Grande, C.A.~Angell, and P.F.~McMillan,
		Science {\bf 275}, 322 (1997).

\bibitem{Harrington-PRL-1997}
		S.~Harrington, R.~Zhang, P.H.~Poole, F.~Sciortino, and
		H.E.~Stanley,
		Phys.\ Rev.\ Lett.\ {\bf 78}, 2409 (1997).

\bibitem{Tse-Nature-1999}
                J.S.~Tse, D.D.~Klug, C.A.~Tulk,
		I.~Swainson, E.C.~Svensson, C.-K.~Loong,
		V.~Shpakov, V.R.~Belosludov, R.V.~Belosludov,
		Y.~Kawazoe, Nature {\bf 400}, 647 (1999).

\bibitem{Fischer-PRL-1999}
		H.E.~Fischer, F.J.~Bermejo, G.J.~Cuello,
 		M.T.~Fern\'andez-Diaz, J.~Dawidowski, M.A.~Gonz\'alez,
 		H.~Schober, and M.~Jimenez-Ruiz,
 		Phys.\ Rev.\ Lett.\ {\bf 82}, 1193 (1999).

\bibitem{tunneling}
		P.W.~Anderson, B.I.~Halperin, and C.M.~Varma,
		Philos. Mag. {\bf 25}, 1 (1972).

\bibitem{relaxational}
		U.~Balucani and M.~Zoppi,
		{\it Dynamics of the Liquid State},
 		(Clarendon Press, Oxford, 1994).

\bibitem{Sette-Science-1998}
		F.~Sette, M.H.~Krisch, C.~Masciovecchio, G.~Ruocco,
		and G.~Monaco, Science {\bf 280}, 1550 (1998).


\bibitem{expdet}
		G.~Ruocco and F.~Sette, J.\ Phys: Cond.\ Matt.\ {\bf 11},
		R259 (1999),
		and references therein.

\bibitem{Mishima-Nature-1984}
                 O.~Mishima, L.D.~Calvert and E.~Whalley, {\it Nature},
                 {\bf 310}, 393 (1984).

\bibitem{Koza-PRL-1999}
		M.~Koza, H.~Schober, A.~T\"olle, Th.~Hanse, and F.~Fujara,
		submitted

\bibitem{Bellissent-JCP-1987}
                M.-C.~Bellissent-Funel, J.~Teixeira,  and L.~Bosio,
                J.\ Chem.\ Phys. {\bf 87}, 2231 (1987).

\bibitem{Koza-Nature-1999}
 		M.~Koza, H.~Schober, A.~T\"olle, F.~Fujara,
		and T.~Hansen,
 		Nature {\bf 397}, 660 (1999).


\bibitem{Schober-PB-1998}
		H.~Schober, M.~Koza. A.~T\"olle, F.~Fujara, C.A.~Angell,
		and R.~B\"ohmer,
		Physica B {\bf 241-243}, 897 (1998).

\bibitem{Caprion-PRL-1996} D.~Caprion, P.~Jund, and R.~Jullien,
		Phys.\ Rev.\ Lett.\ {\bf 77}, 675 (1996).

\bibitem{Angell}
		C.A.~Angell, in {\it "Relaxation in
		Complex Systems"} edited by K.~L.~Ngai
		and G.B.~Wright, NRL p. 3, (Washington, 1984).

\bibitem{Ito-Nature-1999} K.~Ito, C.T.~Moynihan, and C.A.~Angell,
                     Nature {\bf 398}, 492 (1999).

\bibitem{Starr-2000} F.W.~Starr, C.A.~Angell, R.J.~Speedy, and
		H.E.~Stanley, arXiv.cond-mat/9903451.

\bibitem{footnote2} Cubic ice is not a perfect crystal due to the
                    presence of staking faults (W.F.~Kuhs, D.V.~Bliss and
                    J.L.~Finney, J.\ Physique {\bf 48}, C 1631 (1987)).
		    This is an interesting subject in its own, however,
		    without incidence on the discussion of the amorphous phases
		    presented here.

\bibitem{footnote1} It should be pointed out that while ice I$_{c}$ is
                    a good reference system for $LDA$ this may not be the
		    case for $HDA$.
                    $HDA$ has supposedly a bonding topology closer
                    related to high density ice forms like ice XII, which
                    is formed
                    under similar conditions \cite{Koza-Nature-1999}.

\bibitem{Buchenau-ZP-1985} U.~Buchenau, Zeitsch.\ Phys.\ B {\bf 58}, 181
(1985).

\bibitem{Bellissent-EPL-1998}
                    M.-C.~Bellissent-Funel, Europhys.\ Lett.\ {\bf 42}, 161
		    (1998).
\bibitem{Pusztai-PRB-2000}
                    L.~Pusztai, Phys.\ Rev.\ B {\bf 61}, 28 (2000).

\bibitem{Soper-JCP-1994}
             	    A.K.~Soper, J.\ Chem.\ Phys.\ {\bf 101}, 6888 (1994).

\bibitem{Trachenko-PRL-1998}
                    K.~Trachenko, M.T.~Dove, K.D.~Hammonds,
                    M.J.~Harris, and V.~Heine,
		    Phys.\ Rev. Lett.\ {\bf 81}, 3431 (1998).

\bibitem{Taraskin-PRB-1999}
                    S.N.~Taraskin and S.R.~Elliott, Phys.\ Rev.\ B
                    {\bf 59}, 8572 (1999).
\end{references}
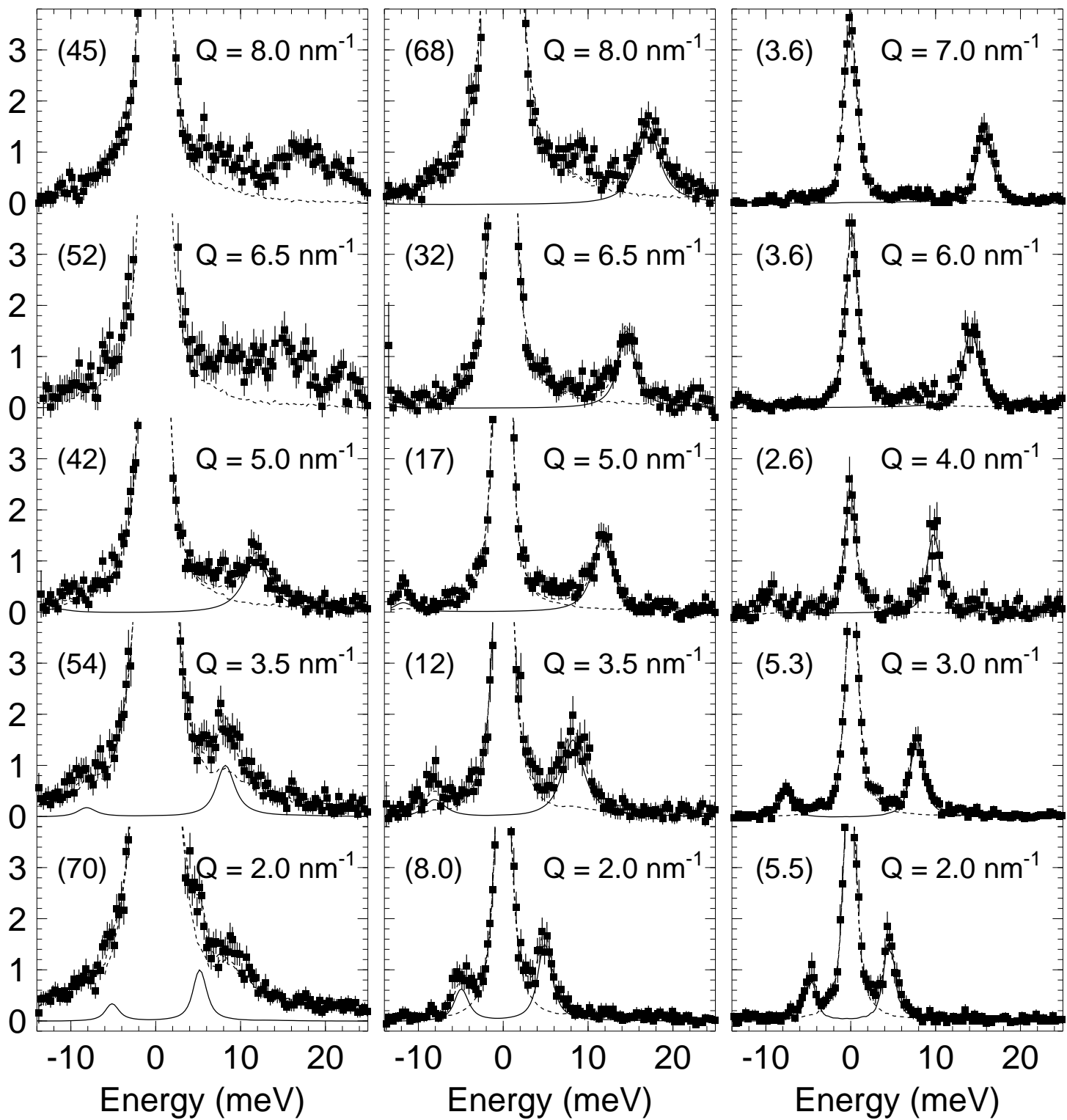
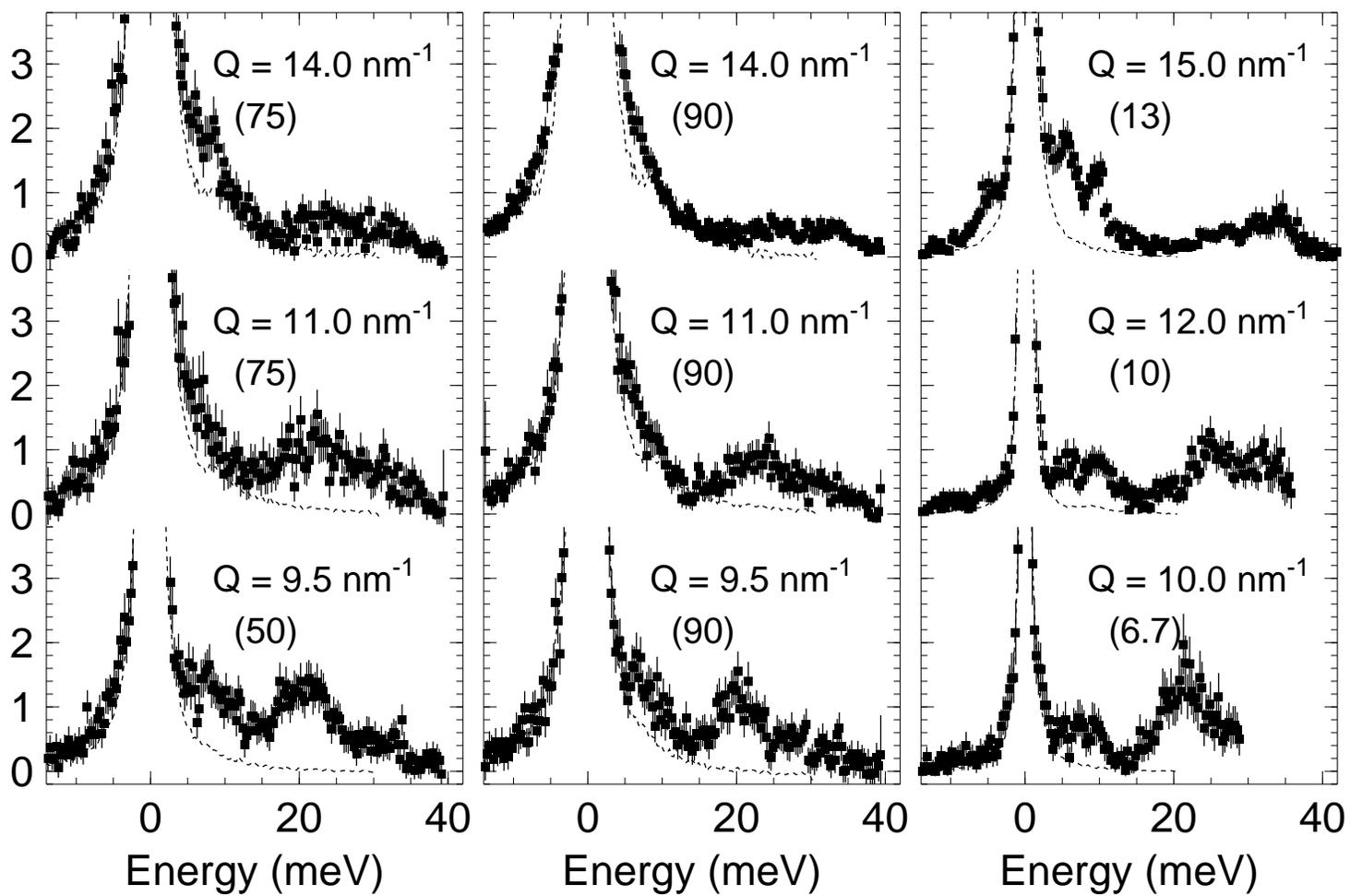
\end{document}